# Universality in coalescence of polymeric fluids

Sarath Chandra Varma[1], Siddhartha Mukherjee[2,3], Aniruddha Saha[2], Aditya Bandopadhyay[2] Aloke Kumar[1*], Suman Chakraborty[2,3*]

[1]Department of Mechanical Engineering, Indian Institute of Science, Bangalore, India

[2]Department of Mechanical Engineering, Indian Institute of Technology, Kharagpur, India

[3]Advanced Technology Development Center, Indian Institute of Technology, Kharagpur, 721302, India

*alokekumar@iisc.ac.in
*suman@mech.iitkgp.ernet.in

## Abstract

Coalescence of liquid droplets involves an interplay between capillary forces, viscous forces and inertial forces. Here, we unveil a universal temporal evolution of the neck radius during the coalescence of two polymeric drops. Through high speed imaging we demonstrate that drops of polyacrylamide (PAM), poly-vinyl alcohol (PVA), polyethylene oxide (PEO), polyethylene glycol (PEG) and xanthan gum (XG) depict a universal behavior $\frac{R}{\sqrt{v_o \lambda}} \propto \left(\frac{t}{\lambda}\right)^{0.36} \left(\frac{c}{c^*}\right)^{-0.83}$ over a dilute, semi-dilute and non-dilute range of concentrations. A linear Phan-Thein-Tanner viscoelastic model captures the temporal aspect of universality.

Coalescence of drops is ubiquitous in natural phenomena such as rain drop condensation[1-4], and industrial phenomena such as spraying[5], coating[6] and even processes linked to life itself[7-10]. Studies on droplet coalescence have focused on the behaviour of Newtonian fluid droplets[11-15], where it has been shown that the kinematics of the phenomenon is governed by the growth of the liquid bridge that can be typically characterized by the temporal change of a scalar variable—the neck radius R(t). For Newtonian fluids the coalescence phenomena is dominated by two universal regimes i) a viscous dominated regime at smaller time-scales[16] and ii) a inertia dominated regime at longer time-scales[17]. Whether such universality can be extended to non-Newtonian fluids remains an open question.

Polymeric fluids, which are a distinct sub-set of viscoelastic fluids, contain macromolecules that can exhibit strongly non-Newtonian characteristics such as polymer chain entanglements and molecular relaxations[18]. Solutions of polymers in a good solvents such as water, can admit various states of polymer chain interactions depending on the level of dilution of the polymer. The co-existence of viscoelastic relaxations along with dissipative and inertial dynamics introduces additional complexity so that the constitutive relations may or may not admit universal solutions. However, for Newtonian fluids, universality in droplet coalescence has been demonstrated both experimentally and analytically[15-17,19-28]. For example, in the viscous dissipation dominated regime the neck radius evolution has been shown to have a $R^* \propto t^* \ln(t^*)$, where with $R^*$ (neck radius non dimensionalized by initial radius of drop $R_o$) and $t^*$ (non dimensionalized time with $\tau_v = \frac{\eta R_o}{\sigma}$ where $\eta$ is viscosity and $\sigma$ is surface tension) proposed by Hopper[23,24,29,30]. In the inertial dominated regime for Newtonian fluids, it has been shown that the radius scales as $R^* \propto (t^*)^{\frac{1}{2}}$ where $\rho$ is density, $t^*$ (non dimensionalized time with by inertial time scale $\tau_i = \frac{\rho R_o^3}{\sigma}$ Eggers et al[22].

In this work, we study the coalescence of droplets of five different water-soluble polymers Polyacramalide (PAM), Polyethylene oxide (PEO), Xanthan gum (XG), Polyvinyl alcohol (PVA) and Polyethylene glycol (PEG). By investigating coalescence phenomena for aqueous solutions of the polymers, at various concentrations varying from dilute to entangled regions, we unveil a universal temporal evolution of the necking radius of polymeric fluid drops. Experimental observation of a universal regime for these polymeric fluids is also supported by scaling analysis based on a linear Phan-Thien-Tanner (PTT) model. Our results stand in strong contrast to the universality found in Newtonian fluids and they hold enormous promise for rheological measurements.

Coalescence of liquid drops proceeds via the formation of liquid bridge, where the neck radius (R) varies with time. Such variation of the bridge for the coalescence of aqueous solution of PAM (0.5% w/v) at two time instants is shown in Fig.1a-b and the variation of neck radius through out the process of coalescence of the chosen polymeric liquids of 0.5% w/v concentration is represented in Fig.1c. In our study, we considered the linear regime of the temporal variation of neck radius to probe the coalescence phenomenon. The fluids that were chosen for this study are polymeric fluids, which can be expected to show shear-thinning behaviour[18]. Fig.2a shows the variation of viscosity $(\eta)$ as a function of shear-rate $(\dot{\gamma})$ for various dilutions of PAM solutions with concentration ratio is concentration of the solution and $\frac{c}{c^*} > 1$ ($c$ is concentration of solution and $c^*$ is critical concentration of polymer); these solutions which are in semi dilute and entangled regimes showed a strong shear-thinning behaviour. The non-Newtonian behaviour of these polymers can be characterized by a relaxation time-scale, $(\lambda)$ which was calculated using Zimm's model[31]. The rheological data provides the infinite shear viscosity $(\eta_\infty)$ and Fig.2a (inset) representing the rheological data in the form of Carreau-Yasuda model[18]. Subsequently, the zero-shear kinematic viscosities $(v_o)$ of PAM at different concentration solution were obtained. Rheological data of PEO, XG and PEG showed similar trend as PAM's and were the relaxation time-scales and zero-shear viscosities were calculated in an analogous manner. Solutions of PVA, which are in dilute regime $\left(\frac{c}{c^*} < 1\right)$ display a constant viscosity with shear-rate and is shown in Fig. 2b. In such cases, extrapolation was used to calculate $\lambda$ and $v_o$.

For droplet coalescence experiments, similarity in trends of temporal variation of neck radius suggested a universal scaling; in order to determine the functional form the neck radius was converted to a non-dimensional neck radius $R^*$ (ratio of neck radius and diffusion length scale $\sqrt{v_o \lambda}$). It is observed that the non-dimensional neck radius has a functional form of

$R^* = f\left(\dfrac{t}{\lambda}, \dfrac{c}{c^*}\right)$ collapsed into a universal scaling as depicted in Fig.3a, when the non-dimensional radius was scaled with non-dimensional time $t^* = \left(\dfrac{t}{\lambda}\right)\left(\dfrac{c}{c^*}\right)^{-2.28}$ ($t$ is time). This is valid for all the five polymer solutions, with concentrations in either the dilute, semi-dilute and non-dilute regimes. Figure 3b and c, show the individual data for PAM and PEG and it is noteworthy that the scaling relationship remains valid over several decades of non-dimensional time $\left(t^*\right)$.

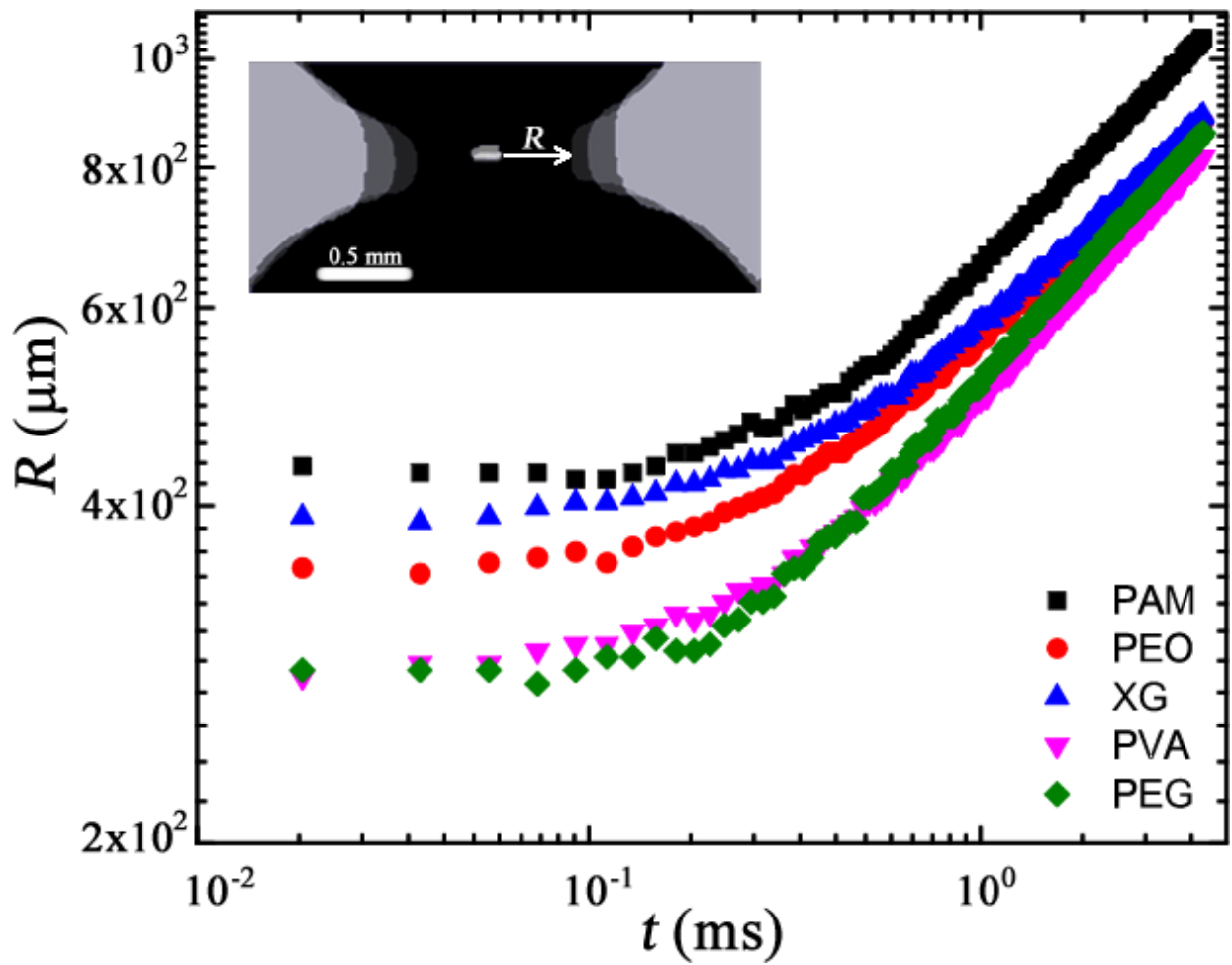

**Figure 1. | Temporal change in neck geometry.** Increase in neck radius with time during merging at 0.5% w/v concentration of polymer solutions used with an inset showing neck radius evolution with time.

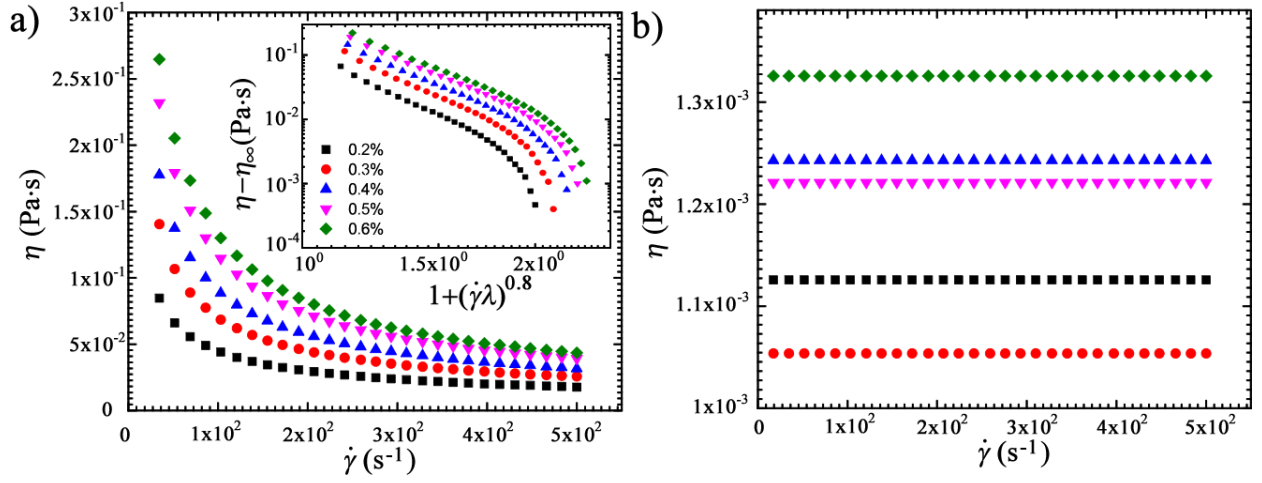

**Figure 2. | Rheological behaviour of solutions**. a) PAM with Inset: Carreau-Yasuda model fitting of rheology data and b) PVA.

Numerically, the scaling law proposed above was verified by considering the special case of $\frac{c}{c^*}=1$, and by using the linear Phan-Thein-Tanner (PTT) model. By assuming steady, unidirectional and axisymmetric flow, the components of the radial-direction momentum terms were scaled using, $\tau \to \bar{\tau} \text{Wi}; P \to \bar{P}\text{Wi}; U \to \bar{U}\sqrt{\text{Wi}}$ ( $\tau$ is stress, $P$ is pressure, $U$ is velocity of neck radius and the quantities represented with bars are scaled parameters and Wi is the Weissenberg number). The scaled stresses were calculated using the approximations, $\lambda \text{Wi}^2 \ll 1$ and $1+\frac{\kappa \lambda \text{Wi}}{\eta}\bar{\tau} \sim \frac{\kappa \lambda \text{Wi}}{\eta}\bar{\tau}$ ($\kappa$ is a constant), which yields

$$\tau_{rr} \sim \sqrt{\frac{2\eta^2}{\kappa \lambda}}\sqrt{\frac{\partial v_r}{\partial r}} \qquad (1)$$

$$\tau_{rz} \sim \tau_{rr}\frac{\left(\frac{\partial v_r}{\partial z}\right)}{2\left(\frac{\partial v_r}{\partial r}\right)} \qquad (2)$$

The following scaling arguments were introduced: $v_r \sim U, r \sim R$ ($R$ is length scale associated with neck geometry), $z \sim R$. The pressure gradient term was modified as $\Delta P = \sigma\left(\frac{1}{H}+\frac{1}{R}-\frac{2}{R_o}\right)$. Further, as shown in Fig. 4, the following relationships were used for quantifying the pressure $P_1 - P_\infty = \frac{2\sigma}{R_o}$, $P_2 - P_\infty = \sigma\left(\frac{1}{H}+\frac{1}{R}\right)$, where $P_1$, $P_2$ are pressures at inside and outside locations of the section and $P_\infty$ is atmospheric pressure. From the geometry of the coalescence process, the relation $\frac{H}{R} \approx \frac{R}{2R_o}$ can be obtained (Fig.4), where H is length scale associated with neck geometry. Finally, the momentum equation can be further simplified by introducing the non-dimensional parameters $R^* = \frac{R}{\sqrt{v_o \lambda}}$ and $t^* = \frac{t}{\lambda}$, which yields:

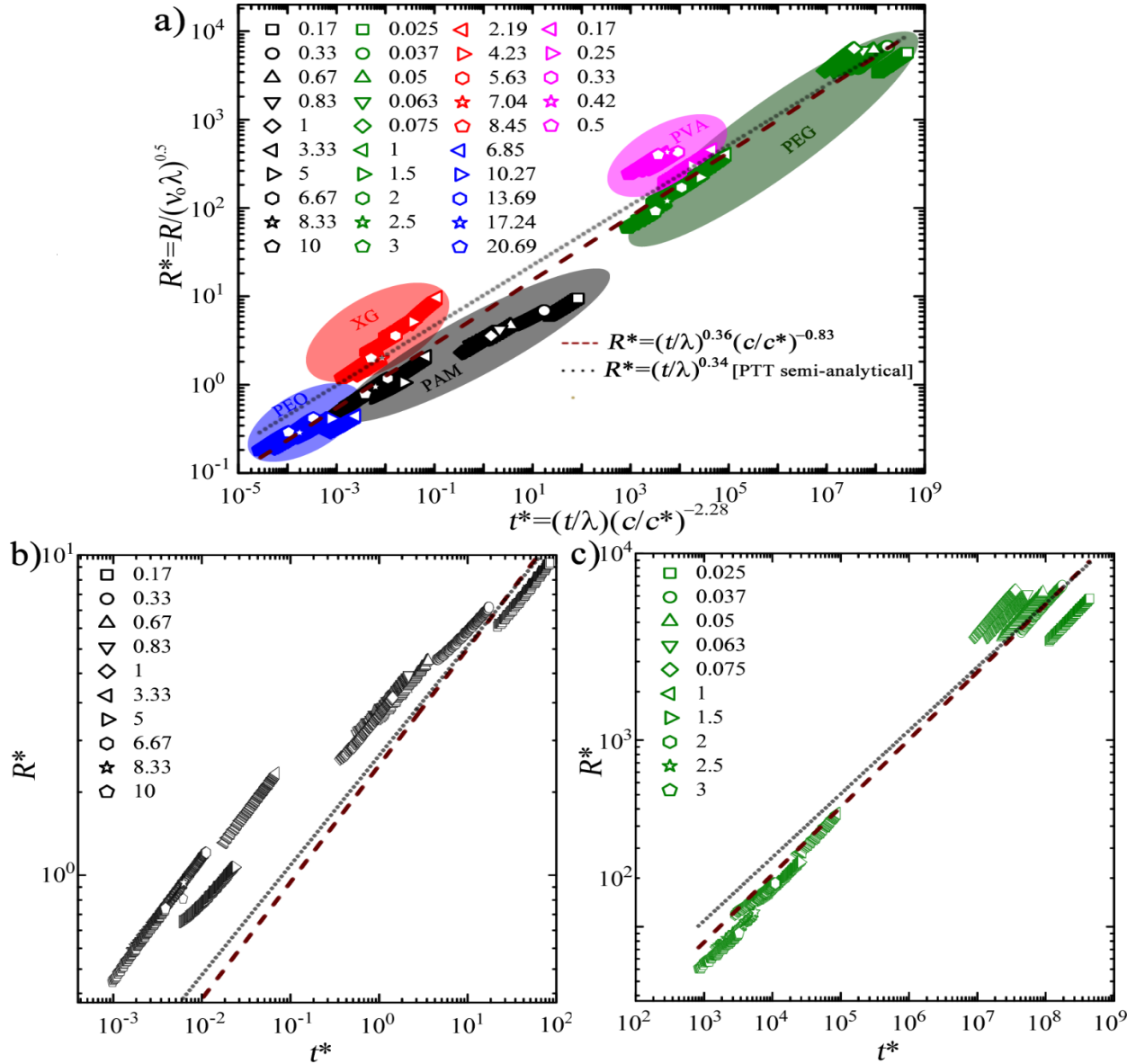

**Figure 3. | Scaling of temporal variation of neck radius.** Neck radius non-dimensionalized with diffusion length scale as function of non-dimensional time and concentration ratios for a) All solutions b) PAM solutions in dilute and semi dilute regimes c) PEG solutions in dilute and semi dilute regimes with legend representing the values of $\frac{c}{c^*}$.

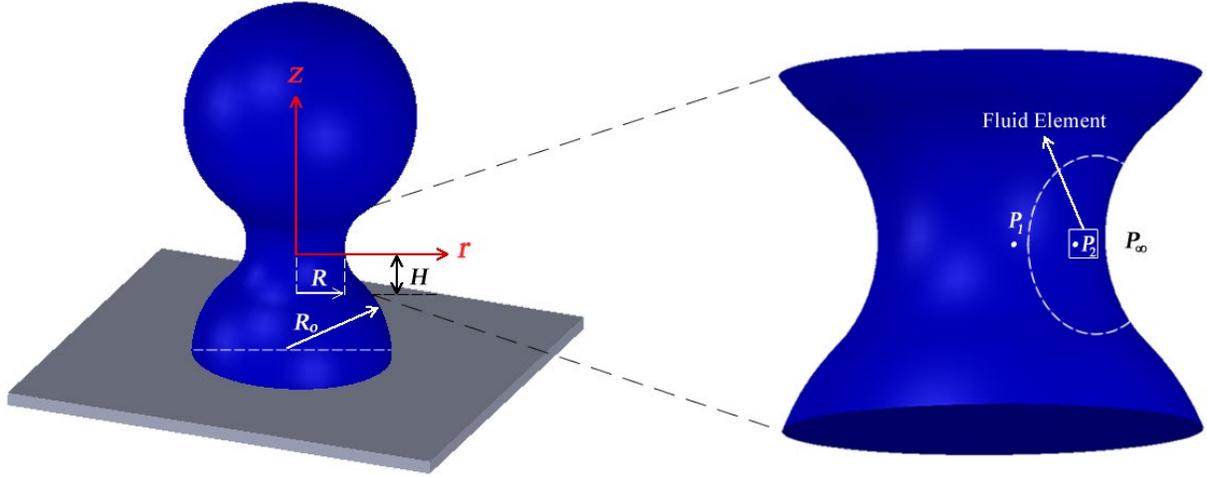

**Figure 4. | Schematic of neck region.** Representation of geometric parameters associated with neck during coalescence.

$$\left(\frac{dR^*}{dt^*}\right)^2 + \frac{C_1}{\sqrt{R^*}}\left(\frac{dR^*}{dt^*}\right)^{\frac{1}{2}} - \frac{C_2}{R^{*2}} = 0 \quad (3)$$

Where, $C_1 \sim \frac{\sqrt{2}\eta}{\rho\sqrt{\kappa\lambda v_o}}$, $C_2 \sim \frac{2\sigma R_o}{\rho v_o \sqrt{\lambda v_o}}$. The numerical solution of (3) leads to the functional form of $R^* \propto (t^*)^{0.34}$, which is represented in Fig.3a along with the experimental result of $R^* \propto (t^*)^{0.36}$. In Fig. 3a, the semi-analytically obtained expression is also 5 compared with the data and experimentally obtained correlation. The experimentally obtained correlation shows good agreement with the semi-analytical model.

In summary we have experimentally and analytically investigated the coalescence of polymeric liquid drops. It has been observed that the scaling laws proposed by previous studies of drop coalescence for Newtonian fluid are not valid in coalescence of viscoelastic liquids. We showed that the variation of non-dimensional neck radius depends on non-dimensional time, where both the relaxation time scale and concentration of the polymer play a role. Finally, we proposed a universal scaling for neck radius which depends on viscoelastic properties of the polymer solutions.